\documentclass[prl,showpacs,nofootinbib,twocolumn]{revtex4}
\usepackage{bm}
\usepackage{stmaryrd}
\usepackage{amsmath}
\usepackage{amssymb}
\usepackage{graphicx}
\usepackage{textcomp}
\usepackage{calrsfs}
\usepackage{yfonts}
\usepackage{epsfig}
\usepackage{pstricks, pst-node, pst-text, pst-3d,pst-coil, epsfig,rotating}
\usepackage{amsmath,graphicx}
\global\arraycolsep=2pt

\newcommand{\ben}{\begin{equation*}}
\newcommand{\een}{\end{equation*}}
\newcommand{\bean}{\begin{eqnarray*}}
\newcommand{\eean}{\end{eqnarray*}}

\newcommand{\nn}{\nonumber}
\newcommand{\be}{\begin{equation}}
\newcommand{\ee}{\end{equation}}
\newcommand{\bea}{\begin{eqnarray}}     
\newcommand{\eea}{\end{eqnarray}}

\begin{document}
\title{Electromagnetic Angular Momentum and Relativity}

\author{Kimball A. Milton}\email{milton@nhn.ou.edu}

\affiliation{
H. L. Dodge Department of Physics and Astronomy, University of Oklahoma,
Norman, OK 73019}
\author{Giulio Meille}\email{giulioemeille@gmail.com}
\affiliation{Department of Physics, Columbia University, New York, NY 10027}

\begin{abstract}Recently there have been suggestions that the Lorentz force law
is inconsistent with special relativity.  This is difficult to understand,
since Einstein invented relativity in order to reconcile electrodynamics
with mechanics.  Here we investigate the momentum of an electric charge
and a magnetic dipole in the frame in which both are at rest, and in
an infinitesimally boosted frame in which both have a common velocity.
We show that for a dipole composed of a magnetic monopole-antimonopole pair
the torque is zero in both frames, while if the dipole is a point dipole,
the torque is not zero, but is balanced by the rate of change of the
angular momentum of the electromagnetic field, so there is no mechanical
torque on the dipole.
\end{abstract}
\date\today

\maketitle

\section{Introduction}
\label{sec1}
Considerable press was given to a recent claim by Mansuripur \cite{mansuripur}
that the Lorentz force law appears to be in contradiction with special
relativity.  Because that force law is based on Maxwell's equations, which
are the origin and basis for Einsteinian relativity, it is difficult to
understand this claim.  Mansuripur's analysis, based on 
Ref.~\cite{mansuripur11},
suggests that the Lorentz force law must be replaced by some other expression,
such as that of Einstein and Laub \cite{el,el2}.  This conclusion has
received widespread publicity in the popular press \cite{science}

By now there have appeared many critiques of this conclusion
\cite{unnik,vanzella,griffiths,cross,mcdonald,boyer,brachet}.  The consensus
seems firm that omitted contributions remove the contradiction. 
However, these analyses seem to consider special cases and somewhat
obscure concepts, such as hidden momentum \cite{furry,griffiths09}, so
it seems to us that a simple general analysis will clarify all points and
settle the issue definitively.

The problem posed by Mansuripur is the following.  Consider an electric
charge and a magnetic dipole separated by some distance at rest.  Evidently,
there is no force or torque on either particle.  When the same system is
viewed in a boosted frame, where both particles have a common velocity,
it appears that the torque on the dipole
\be
\bm{\tau}=\bm{\mu}\times \mathbf{B}+\mathbf{d\times E}\label{naivetorque}
\ee
is nonzero.  Thus in this frame it seems there should be precession of
the dipole, violating the principle of relativity.

In this note we revisit this problem in the following way.  %In Sec.~\ref{sec2}
We first compute the momentum and angular momentum of the electromagnetic field
for the static configuration of charge and dipole; the values of these quantities
are different for a point dipole and a dipole composed of a magnetic 
monopole-antimonopole pair.  This bears on the issue of ``hidden momentum.''
%In Sec.~\ref{sec3}
Then we calculate the charge and current densities, and the fields,
in the boosted frame, and directly compute the torque on the dipole from the microscopic
Lorentz force law.  This torque identically vanishes for a dipole composed of
a magnetic monopole-antimonopole, but is nonzero for a point dipole.  However,
the latter is precisely cancelled by the nonzero rate of change of the
angular momentum of the electromagnetic field in this case, so in either event,
there is no mechanical torque on the dipole.

\section{Momentum and angular momentum of charge and dipole}\label{sec2}
In this section we consider a static system consisting of a charge $e$, which
we may take at the origin, and a dipole $\bm{\mu}$ at position $\mathbf{R}$,
in vacuum.
We will calculate the field momentum and angular momentum for this 
configuration from the formulas (throughout this paper we use Gaussian units;
our primary reference is Ref.~\cite{embook})
\begin{subequations}
\bea
\mathbf{P}&=&\int (d\mathbf{r})\,\mathbf{G},\quad \mathbf{G}=
\frac{\mathbf{E\times B}}{4\pi c},\label{p}\\
\mathbf{J}&=&\int(d\mathbf{r})\,\mathbf{r\times G}.\label{j}
\eea
\end{subequations}

Let us first adopt the model (``Gilbert'') in which
the magnetic dipole consists of a magnetic monopole of charge $+g$ separated
by a displacement $\mathbf{a}$ from a monopole of charge $-g$.  We will always
suppose that $a\ll R$.  The momentum of a system consisting of one electric
charge $e$ and one magnetic pole $g$ is
\be
\mathbf{P}_{eg}=\frac{eg}{4\pi c}\int (d\mathbf{r})\frac{\mathbf{r}}{r^3}
\times\frac{\mathbf{r-R}}{|\mathbf{r-R}|^3}.
\ee
Now the following dyadic (and convergent) integral is easily worked out:
\be
\int (d\mathbf{r})\frac{\mathbf{r}}{r^3}
\frac{\mathbf{r-R}}{|\mathbf{r-R}|^3} =\frac{2\pi}R\left(\bm{1}-\frac{
\mathbf{RR}}{R^2}\right),\label{int}
\ee
from which we immediately conclude that the momentum $\mathbf{P}_{eg}=0$.
From this it follows that the momentum of the charge and a Gilbert dipole
is also zero,
\be
\mathbf{P}_{eg\bar g}=0.\label{mommu}
\ee

The same integral (\ref{int}) allows us to conclude that the angular
momentum of a charge and a magnetic monopole is nonzero,
\be
\mathbf{J}_{eg}=-\frac{eg}{4\pi c}\int(d\mathbf{r})\mathbf{r}\times
\frac{\mathbf{r\times R}}{r^3|\mathbf{r-R}|^3}=\frac{eg}c \frac{\mathbf{R}}R,
\ee
a result first discovered by Poincar\'e \cite{poincare} and then by Thomson
\cite{thomson}.  From this the angular momentum for a dipole made from a
monopole-antimonopole separated by a distance $\mathbf{a}$ is
\be
\mathbf{J}_{eg\bar g}=\frac{eg}c\mathbf{a}\cdot \bm{\nabla}\frac{\mathbf{R}}R=\frac1c\mathbf{R}
\times(\bm{\mu}\times\mathbf{E(R)}),\label{angmomemu}
\ee
where $\bm{\mu}=g\mathbf{a}$ and $\mathbf{E}=e\mathbf{R}/R^3$.

Now consider a point magnetic dipole, which possesses the magnetic field
\be
\mathbf{B}=\mathbf{B}_s+\mathbf{B}_f,\label{bmu}
\ee 
where the second term is the usual field of a magnetic dipole,
\begin{subequations}
\bea
\mathbf{B}_{f}&=&-\bm{\nabla}\left(\bm{\mu}
\cdot\frac{\mathbf{r-R}}{|\mathbf{r-R}|^3}\right)\nn\\
&=&\frac{3(\mathbf{r-R})\,\bm{\mu}\cdot(\mathbf{r-R})-(\mathbf{r-R})^2
\bm{\mu}}{|\mathbf{r-R}|^5}.
\eea
The first term in Eq.~(\ref{bmu}) is required to satify the magnetic-charge-free Maxwell
equation $\bm{\nabla}\cdot\mathbf{B}=0$:
\be
\mathbf{B}_s=4\pi\bm{\mu}\delta(\mathbf{r-R}).
\ee
\end{subequations}
If we ignore the latter for the moment,
 it is straightforward to work out the angular momentum from the
field part,
\bea
\mathbf{J}_{fe\mu}&=&\int(d\mathbf{r})\mathbf{r}\times\frac{e}{4\pi c}
\left(\frac{\mathbf{r}}{r^3}\times\bm{\nabla}_{\mathbf R}\right)
\frac{\bm{\mu}\cdot (\mathbf{r-R})}{|\mathbf{r-R}|^3}\nn\\
&=&\frac{e}{4\pi c}\bm{\mu}\cdot\bm{\nabla}_{\mathbf{R}}
\bm{\nabla}_{\mathbf{R}}\cdot\int(d\mathbf{r})\left(\frac{\mathbf{r r}}{r^3}-\frac{\bm{1}}r
\right)\frac1{|\mathbf{r-R}|}.\nn\\
\eea
The latter integral is evaluated to be $\pi R(\bm{1}+\mathbf{RR}/R^2)$
and then the same result (\ref{angmomemu}) follows.

However, for a point dipole, there is another contribution from the
$\delta$-function term in Eq.~(\ref{bmu}) which exactly cancels the
above field part,
\be
\mathbf{J}_{e\mu}=0. \label{angmomptmu}
\ee
This term further gives a contribution to the field momentum:
\be
\mathbf{P}_{e\mu}=-\frac1c\bm{\mu}\times \mathbf{E}(\mathbf{R}). \label{momptmu}
\ee
These results for the Gilbert dipole, Eqs.~(\ref{angmomemu}) and (\ref{mommu}),
 and for the Amp\`ere dipole, Eqs.~(\ref{angmomptmu}) and (\ref{momptmu}),
agree partly with those found by Furry \cite{furry}, but he asserts without
evidence that the Amp\`ere angular momentum is unaltered from the Gilbert one,
whereas his linear momentum seems to agree with ours.

\section{Boosted frame}\label{sec3}
Our system, in the rest frame, in the Gilbert description, is defined
by the electric and magnetic charge densities,
\begin{subequations}
\bea
\rho_e&=&e\delta(\mathbf{r}),\quad \mathbf{j}_e=0,\\
\rho_m&=&-\bm{\mu}\cdot\bm{\nabla}\delta(\mathbf{r-R}),\quad \mathbf{j}_m=0.
\eea
\end{subequations}
Now consider a infinitesimally boosted frame, in which all particles
are moving with velocity $\delta\mathbf{v}\ll c$.
In such a frame the electric and magnetic fields are modified 
(Ref.~\cite{embook}, Sec.~10.3)
\begin{subequations}
\bea
\delta \mathbf{E}&=&-\delta_{\rm coor}\mathbf{E}-\frac1c\delta\mathbf{v
\times B},\\
\delta \mathbf{B}&=&-\delta_{\rm coor}\mathbf{B}+\frac1c\delta\mathbf{v
\times E},
\eea
\end{subequations}
where 
\be
\delta_{\rm coor}=\delta \mathbf{v}t\cdot\bm{\nabla}+\frac1{c^2}\delta
\mathbf{v\cdot r}\frac\partial{\partial t}.
\ee
Then from Maxwell's equations with magnetic charge
\begin{subequations}
\bea
\bm{\nabla}\cdot \mathbf{E}&=&4\pi \rho_e,\quad -\bm{\nabla} \times \mathbf{E}=
\frac1c\frac{\partial}{\partial t}\mathbf{B}+\frac{4\pi}c\mathbf{j}_m,\\
\bm{\nabla}\cdot \mathbf{B}&=&4\pi \rho_m,\quad \bm{\nabla} \times \mathbf{B}=
\frac1c\frac{\partial}{\partial t}\mathbf{E}+\frac{4\pi}c\mathbf{j}_e,
\eea
\end{subequations}
we deduce
\begin{subequations}
\bea
\delta\rho_e&=&-\delta_{\rm coor}\rho_e+\frac1{c^2}\delta\mathbf{v\cdot j}_e,\\
\delta\mathbf{j}_e&=&-\delta_{\rm coor}\mathbf{j}_e+\delta\mathbf{v}\rho_e,\\
\delta\rho_m&=&-\delta_{\rm coor}\rho_m+\frac1{c^2}\delta\mathbf{v\cdot j}_m,\\
\delta\mathbf{j}_m&=&-\delta_{\rm coor}\mathbf{j}_m+\delta\mathbf{v}\rho_m,\nn\\
\eea
\end{subequations}

To compute the torque in the boosted frame, we start from first principles,
and use the microscopic Lorentz force density,
\be
\mathbf{f}=\rho_e\mathbf{E}+\frac1c\mathbf{j}_e\times\mathbf{B}+
\rho_m\mathbf{B}-\frac1c\mathbf{j}_m\times\mathbf{E}.
\ee
(Note the dual symmetry, replacing electric quantities by magnetic ones, and 
magnetic quantities by the negative of electric ones.)
In the present problem, there is an induced magnetic charge density
and current density,
%\begin{subequations}
\be
\delta \rho_m=-\delta\mathbf{v}t\cdot\bm{\nabla}\rho_m,\quad
\delta\mathbf{j}_m=\delta\mathbf{v}\rho_m,
\ee
as well as an induced magnetic field acting on the dipole,
\be
\delta \mathbf{B}=\frac1c\delta\mathbf{v\times E}.
\ee
Thus the torque on the dipole in the moving frame is
\bea
\bm\tau&=&\int (d\mathbf{r})\,\mathbf{r}\times\left(\rho_m\delta \mathbf{B}
-\frac1c\delta\mathbf{j}_m\times \mathbf{E}\right)\nonumber\\
&=&\int (d\mathbf{r})\,\mathbf{r}\times\left(\rho_m\frac{\delta\mathbf{v}}c
\times\mathbf{E}-\frac{\delta \mathbf{v}}c\rho_m\times \mathbf{E}\right)=0,
%&=&\bm{\mu}\times\left(\frac{\delta\mathbf{v}}c\times\mathbf{E}\right)
%+\mathbf{R}\times\left(\frac{\delta\mathbf{v}}c\times(\bm{\mu}\cdot\bm{\nabla})
%\mathbf{E}\right)\nonumber\\
%&&-\bm{\mu}\times\left(\frac{\delta\mathbf{v}}c\times\mathbf{E}\right)
%-\mathbf{R}\times\left(\frac{\delta\mathbf{v}}c\times(\bm{\mu}\cdot\bm{\nabla})
%\mathbf{E}\right),\nn\\
\eea
that is, the two terms exactly cancel.  In precisely the same way, we
can show that the torque on the electric charge vanishes as well.
Thus there is no torque and no violation of relativistic invariance.
The reason for the apparent anomaly was the invalid use of Eq.~(\ref{naivetorque}),
which applies only for constant fields, and cannot be applied indiscriminately.
In fact, a discussion equivalent to this is given in Problem 3, Chapter 4,
in Ref.~\cite{embook}.

So there is no anomaly for a Gilbert dipole.  What happens for an Amp\`erian
one?  In that case we have no magnetic charge or current, so the torque
is
\bea
\bm{\tau}&=&
\int (d\mathbf{r})\mathbf{r}\times\left[\frac{\delta\mathbf{v}}{c^2}
\cdot\mathbf{j}_e\,\mathbf{E}+\frac1c\mathbf{j}_e
\times\left(\frac{\delta \mathbf{v}}c\times\mathbf{E}\right)\right]\nn\\
&=&-\frac{\delta \mathbf{v}}{c^2}\times\int (d\mathbf{r})\mathbf{r\,j\cdot E}.
\eea
Now, ignoring an electric quadrupole term [Chapter 32, Ref.~\cite{embook}],
we may replace $\mathbf{r\,j}\to\frac12(\mathbf{r\,j-j\,r})$, and so we
evaluate
\be
\bm{\tau}=\frac{\delta \mathbf{v}}c\times(\bm{\mu}\times\mathbf{E}),
\ee
which uses the definition of the magnetic moment,
\be
\bm{\mu}=\frac1{2c}\int(d\mathbf{r})\,\mathbf{r\times j(r)}.
\ee
The torque on the dipole does not vanish.

But now there is a contribution from the field angular momentum in
the boosted frame:
\be
\delta \mathbf{J}=\frac1{4\pi c}\int(d\mathbf{r})\,\mathbf{r}\times
\left(\delta\mathbf{E\times B}+\mathbf{E}\times\delta
\mathbf{B}\right).
\ee
 This is simply seen to be
\be
\delta \mathbf{J}=\delta\mathbf{v}t\times \mathbf{P}_{e\mu}+\mbox{constant},
\ee
where the time-dependent term arises from the coordinate variation.
This angular momentum is expressed
in terms of the momentum of the field in the rest frame.  The latter is zero
for a Gilbert dipole, but is given by Eq.~(\ref{momptmu}) for a point dipole.
Thus, in the latter case,
\be
\bm{\tau}+\frac{d}{dt}\delta\mathbf{J}=0,
\ee
So again there is no inconsistency.

\section{Conclusions}
\label{sec5}
In this note we have examined the question of the torque on a magnetic
dipole due to an electric charge.  There is, in constrast to claims
in the literature \cite{mansuripur}, no contradiction with special relativity,
which could hardly be otherwise, since the equations of electromagnetism are
the origin of Einsteinian relativity.  The details of how this consistency
is achieved depends on the model of the dipole.  If we apply the simple fiction that
the dipole consists of a monopole-antimonopole pair, the torque, computed
from the microscopic Lorentz force density, vanishes in the boosted frame.
If the model of the dipole is consistent with the Amp\`erian hypothesis
that all magnetic effects emerge from moving electric charges or changing
electric fields, the torque
is not zero, but is balanced by the change in the field momentum.  So there
is no mechanical torque on the dipole.

Although there has been a chorus of critiques of the claim of Ref.~\cite{mansuripur},
none has seemed definitive to us.  The arguments seem to rely on special models
and approximate expressions.  We have not adopted any specific model, but considered
either the fictional but useful picture of a Gilbert dipole composed of a magnetic 
monopole-antimonopole pair, or that of a generic infinitesimal Amp\`erian current loop.  On that
basis the consistency of the theory, in particular, the validity of the microscopic
Lorentz force law, is established simply from Maxwell's equations without further 
assumptions.
 
\acknowledgments
This work was supported by grants from the US National Science
Foundation and the US Department of Energy.  We thank Elom Abalo, Nathan Beck,
Nathan Edmonsond, Prachi Parashar, K. V. Shajesh,
 and Jimmy Wu for helpful discussions.
GM thanks the NSF-funded REU program at OU for summer support.

\end{document}